\documentclass[12pt]{article}

\def\dspace{\baselineskip = 0.30in}

\input epsf.tex
\def\DESepsf(#1 width #2){\epsfxsize=#2 \epsfbox{#1}}

\begin{document}
\dspace

\begin{titlepage}
\begin{flushright}
April 2002\\
\end{flushright}
\vskip 2cm
\begin{center}
{\Large\bf Four Puzzles of Neutrino Mixing}
\vskip 1cm
{\normalsize\bf
S.M. Barr}
\vskip 0.5cm
{\it Bartol Research Institute, University of Delaware, \\Newark,
DE~~19716\\[0.1truecm]
}

\end{center}
\vskip .5cm

\begin{abstract}

There are four puzzling questions about by the magnitudes
of neutrino mixings and mass splittings. A brief sketch is given of
the various kinds of models of neutrino masses and how they answer
these questions. Special attention is given to so-called ``lopsided"
models.

\end{abstract}
\end{titlepage}

\newpage

\section{Comparison of Quarks and Leptons}
Over the last three decades theorists have been trying to understand
the spectrum of quark and lepton masses. Although no simple model of
the many that have been proposed is uniquely compelling, there are
certain basic ideas that seem rather probable and are
incorporated in most published models. One of these ideas is that
there is a direct relation between the mass ratios and the mixing angles
of the quarks. Since the charged leptons exhibit a mass hierarchy
very similar to those of the quarks, it was widely expected that the
lepton mixing angles would also be like those of the quarks.
The discovery that the atmospheric neutrino mixing angle $\theta_{atm}$
is nearly maximal thus came as a surprise. 

In this talk I first review the basic facts about quark masses and mixings
and then discuss several features of neutrino mixing that seem at first
sight puzzling in light of these facts. I will then show how various types 
of models explain these puzzling features.

There are two quark mass matrices, $M_U$ and $M_D$, for the up-type
quarks ($u$, $c$, $t$) and down-type quarks ($d$, $s$, $b$) respectively.
These are diagonalized by unitary transformations: $V_U^{\dag} M_U U_U
= {\rm diag} (m_u, m_c, m_t)$ and $V_D^{\dag} M_D U_D
= {\rm diag} (m_d, m_s, m_b)$. The mismatch between the unitary transformations
of the left-handed quarks gives rise to the CKM matrix, $V_{CKM} =
U_U^{\dag} U_D$. The CKM angles are $|V_{us}| = \sin \theta^q_{12} \cong 0.2$, 
$|V_{cb}| = \sin \theta^q_{23} \cong 0.04$, and $|V_{ub}| = \sin \theta^q_{13}
\cong 0.003$. The smallness of these angles is presumably due to small
ratios of elements in $M_U$ and $M_D$, and is
therefore presumably directly related to the smallness of the
mass ratios $m_c/m_t$, $m_u/m_c$, $m_s/m_b$, and $m_d/m_s$

How mass ratios and mixing angles might be directly related can be seen
easily from a $2 \times 2$ example \cite{wzf}. Consider the matrix 
\begin{equation}
M = \left( \begin{array}{cc} 0 & \epsilon \\ \epsilon & 1 \end{array}
\right) m.
\end{equation}

\noindent
This is diagonalized by $R(\theta)^T M R(\theta)$, where $R(\theta)$
is the $2 \times 2$ rotation matrix with $\tan 2 \theta = 2 \epsilon$,
or, for small $\epsilon$, $\theta \cong \epsilon$. The large eigenvalue
of $M$ is obviously $m_2 \cong m$, while the fact that $\det M
= - \epsilon^2 m$, tells us that other eigenvalue is $m_1 \cong
- \epsilon^2 m$. Consequently, one has that $\theta \cong \sqrt{|m_1/m_2|}$.
This can be compared to the old and famously successful relation
for the Cabibbo angle $\tan \theta_c \cong \sqrt{m_s/m_d}$.

One should note that the matrix in this example is ``hierarchical", by
which we mean that the entries get smaller {\it upward} and {\it to the
left} of any diagonal entry. Most realistic models of quark
masses and mixings assume such hierarchical mass matrices. For example,
a recent model of Babu and Nandi \cite{bn}, which fits the data extremely well,
has quark matrices of the form
\begin{equation}
M_U \sim \left( \begin{array}{ccc} 
\epsilon^6 & \epsilon^4 & \epsilon^4 \\
\epsilon^4 & \epsilon^2 & \epsilon^2 \\
\epsilon^4 & \epsilon^2 & 1 \end{array} \right) m, \;\;\;
M_D \sim \left( \begin{array}{ccc} 
\epsilon^6 & \epsilon^6 & \epsilon^6 \\
\epsilon^6 & \epsilon^4 & \epsilon^4 \\
\epsilon^6 & \epsilon^4 & \epsilon^2 \end{array} \right) m,
\end{equation}

Now let us consider the leptons. Here again there are two mass matrices,
$M_L$ for the charged leptons ($e^-$, $\mu^-$, $\tau^-$) and $M_{\nu}$
for the neutrinos ($\nu_e$, $\nu_{\mu}$, $\nu_{\tau}$). The matrix $M_{\nu}$
is different in two respects from $M_U$, $M_D$, and $M_L$: it has much
smaller entries, and it is symmetric, since it is a Majorana matrix connecting
left-handed neutrinos to left-handed neutrinos. Nevertheless, as with
the quark mass matrices, the lepton mass matrices are diagonalized by
unitary transformations that can have a mismatch. That mismatch gives rise 
to the neutrino mixing matrix sometimes referred to as the MNS 
matrix \cite{mns}.
$U_{MNS} = U_L^{\dag} U_{\nu}$. In $(U_{MNS})_{fm}$, $f = e,\mu,\tau$
and $m = 1,2,3$. Experimentally one has that $|U_{\mu 3}| (\equiv 
\sin \theta_{atm} = \sin \theta^{\ell}_{23}) \cong 0.7$, $|U_{e2}| (\equiv
\sin \theta_{sol} = \sin \theta^{\ell}_{12}) = O(1)$ (probably), and
$|U_{e3}| (\equiv \sin \theta^{\ell}_{13}) \leq 0.15$. 
There is still a great deal
of uncertainty about the solar mixing angle, but the solution with small
$\theta_{sol}$ (the ``SMA" or Small Mixing Angle MSW solution) is
disfavored by recent global fits to the data \cite{fits}. 
The best fits are to the
``LMA" or Large Mixing Angle MSW solution and the ``LOW" solution.
The best-fit value for the LMA solution is $\tan^2 \theta_{sol} \approx
0.4$. 

The mass splittings needed to fit the atmospheric and solar data are
$\delta m^2_{atm} = m_3^2 - m_2^2 \approx 3 \times 10^{-3}$ eV$^2$, and
$\delta m^2_{sol} = m_2^2 - m_1^2 \approx 5 \times 10^{-5}$ eV$^2$ (for LMA,
smaller for other solar solutions). The fact that $\delta m^2_{sol} \ll
\delta m^2_{atm}$ suggests there is probably a family hierarchy of
neutrino masses, although it also possible that the neutrino masses
are nearly degenerate and that only their splittings have a hierarchy.
\section{Three Puzzles}
In the basic facts about neutrino masses and mixings there are three 
features that appear puzzling in light of the conventional 
wisdom about quark masses and mixings.

{\bf Puzzle 1:} {\it Why are some $\theta^{\ell} \sim 1$ whereas
all $\theta^q \ll 1$ ?} In grand unified theories the quarks and 
leptons are related, and one expects similar mass ratios and mixing 
angles for them. In models with flavor
symmetry the same flavor symmetries generally control the quark and lepton 
mass matrices and give them similar structure. Empirically, one indeed sees
that the charged leptons have a mass hierarchy qualitatively similar
to those of the up-type and down-type quarks. 
Another similarity is that the 13 mixing angle is
by far the smallest in both cases ($|V_{ub}| \ll |V_{us}|, |V_{cb}|$ and
$|U_{e3}| \ll |U_{e2}|, |U_{\mu 3}|$). 

In light of the expected and actual similarities of quarks and leptons
it appears strange that at least one and probably two of the leptonic
angles are large, while all the quark angles are very small.

{\bf Puzzle 2:} {\it How can there be small lepton mass ratios but large
leptonic mixing angles?} As we have seen, for the quarks the smallness 
of the mixing angles and mass ratios are generally thought to be
related. For the charged leptons the mass ratios are certainly small,
and for the neutrinos at least the ratios of mass splittings are small,
and yet the leptons are very strongly mixed.

{\bf Puzzle 3:} {\it Why are two leptonic angles large but the third
($\theta^{\ell}_{13}$) small?} If all the leptonic angles were of order
unity it might suggest that all the entries of the neutrino mass matrix
$M_{\nu}$ were of the same order, as would typically be the case if it
were a ``random" matrix, as has indeed been suggested \cite{random}.
However, such a matrix
would not generally give a hierarchy of neutrino mass splittings, nor 
would it generally yield a 13 mixing angle much smaller than the others.
The smallness of $\theta^{\ell}_{13}$ and largeness of the other 
leptonic angles suggests that the leptonic mass matrices have quite
special forms. To see what those forms might be let us consider the following
product of rotation matrices:
\begin{equation}
\begin{array}{ccl}
R_{23}(\theta_{atm}) R_{12}(\theta_{sol}) & = &
\left( \begin{array}{ccc} 1 & 0 & 0 \\ 0 & c_{atm} & s_{atm} \\
0 & -s_{atm} & c_{atm} \end{array} \right) \left( \begin{array}{ccc}
c_{sol} & s_{sol} & 0 \\ -s_{sol} & c_{sol} & 0 \\ 0 & 0 & 1 \end{array}
\right) \\ & = & \left( \begin{array}{ccc} c_{sol} & s_{sol} & 0 \\
-c_{atm} s_{sol} & c_{atm} c_{sol} & s_{atm} \\
s_{atm} s_{sol} & -s_{atm} c_{sol} & c_{atm} \end{array} \right)
\end{array}.
\end{equation}

\noindent
One sees that even if $\theta_{atm}$ and $\theta_{sol}$ are both large
this matrix has the property that the 13 element vanishes.
Thus Puzzle 3 is resolved if one has that $U_{MNS} \equiv U_L^{\dag} U_{\nu}
\cong R_{23}(\theta_{atm}) R_{12}(\theta_{sol})$. There are three simple
possibilities:

\vspace{0.1cm} 

\noindent
{\bf Solution A:} $U_L \cong I$, $U_{\nu} \cong R_{23}(\theta_{atm}) 
R_{12}(\theta_{sol})$. 

\noindent
Both large mixing angles come from $M_{\nu}$, whose diagonalization
involves {\it first} a large 23 rotation and {\it then} a large 12 rotation.
\vspace{0.1cm} 

\vspace{0.1cm}

\noindent
{\bf Solution B:} $U_L \cong R_{12}(\theta_{sol}) R_{23}(\theta_{atm})$,  
$U_{\nu} \cong I$. 

\noindent
Both large mixing angles come from $M_L$, whose diagonalization
involves {\it first} a large 12 rotation and {\it then} a large 23 rotation.

\vspace{0.1cm}

\noindent
{\bf Solution C:} $U_L \cong R_{23}(\theta_{atm})$, $U_{\nu} \cong 
R_{12}(\theta_{sol})$. 

\noindent
The large atmospheric angle comes from $M_L$, and the large solar angle
comes from $M_{\nu}$.
\section{How Non-see-saw Models Resolve the Puzzles}

Let us first recall how the see-saw mechanism works. The up quarks, 
down quarks, and charged leptons all have Dirac masses through the
Higgs doublet field (or fields) coupling the left-handed fermions to their
right-handed partners. If there are right-handed neutrinos, then an
analogous Dirac mass matrix $M_{\nu}^{Dirac}$ can exist for the 
neutrinos as well. However, there can also exist a Majorana mass matrix
$M_R$ connecting the right-handed neutrinos to themselves. These
right-handed Majorana masses can be superlarge as they do not break the gauge
symmetries of the Standard Model. Integrating out the superheavy right-handed
neutrinos leaves behind light left-handed neutrinos with an effective
Majorana mass matrix given by the ``see-saw" formula
$M_{\nu} = - M_{\nu}^{Dirac \; T} M_R^{-1} M_{\nu}^{Dirac}$. In see-saw models,
then, the neutrino masses have fundamentally the same origin as the 
charged lepton and quark masses, namely they come from the existence of
both left- and right-handed components coupled together by the doublet Higgs 
field (or fields).

In non-see-saw models there are no right-handed neutrinos.
The masses of the neutrinos therefore have to arise in some other, completely
new way {\it not} directly related to mass generation for the
charged leptons and quarks. Many such mechanisms have been proposed \cite{bd}.
Three popular ones are the Zee mechanism, R-parity violation in SUSY models,
and triplet Higgs.

In the Zee mechanism \cite{zee}, there exists a singly charged, singlet
scalar field $h^+$, which can couple to a
pair of lepton doublets ($h^+ L_i L_j$) and to a pair of Higgs
doublets ($h^+ H_{\alpha} H_{\beta}$). Obviously,
with both types of couplings, there is no way consistently to assign
lepton number to $h^+$. Whether one assigns it $L= -2$ or $L=0$, one
of its couplings will violate lepton number by two units, which is what is
needed to generate Majorana masses for the left-handed neutrinos.
Such masses arise at one-loop.

In theories with low-energy supersymmetry, the neutrinos can acquire mass
by coupling to a neutralino that plays the
role of right-handed neutrino. The scalar that couples the neutrino to the
neutralino is the sneutrino, which is able to obtain a non-zero
vacuum expectation value if R-parity is violated. R-parity violation 
also allows superpotential terms of the type $L Q D^c$
and $L L E^c$, which give one-loop neutrino masses when the 
sleptons and squarks are integrated out. 

Finally, if there is a triplet higgs field $T$ with Standard Model 
quantum numbers $(1,3, +1)$, then it can have a renormalizable coupling
to a pair of lepton doublets ($T L_i L_j$) that gives a tree-level
neutrino mass.

The great advantage of such non-see-saw mechanisms is that they automatically
provide a very plausible answer to Puzzle 1: the lepton mixing angles differ
so dramatically from the quark mixing angles simply because $M_{\nu}$
has a very different origin than $M_U$ and $M_D$. We will now look at 
specific non-see-saw ideas to see how they resolve the other Puzzles.

{\it Inverted Hierarchy Models.} In inverted hierarchy models the neutrino
mass matrix has approximately the following form:
\begin{equation}
M_{\nu} \cong \left( \begin{array}{ccc} 0 & A & B \\ A & 0 & 0 \\
B & 0 & 0 \end{array} \right),
\end{equation}

\noindent
with $A \sim B$.
This can arise in various ways. In the Zee model the one-loop mass matrix
is symmetric with vanishing diagonal elements. If for some reason
the 23 (32) elements are smaller than the others, the form in Eq. (4) results.
It can also result from an approximate $L_e - L_{\mu}
- L_{\tau}$ symmetry.

One can diagonalize the large elements $A$ and $B$ in Eq. (4) by two 
successive large rotations. First, one can rotate in the ``23 plane" by
angle $\theta_{23} \cong \tan^{-1} (B/A) = O(1)$ to eliminate the 13 and 31
elements. Then one can rotate in the ``12 plane" by $\theta_{12} \cong
\pi/4$ to eliminate the 12 and 21 elements:
\begin{equation}
\left[ \begin{array}{ccc} 0 & A & B \\ A & 0 & 0 \\ B & 0 & 0 \end{array}
\right] \begin{array}{c} \longrightarrow \\ \theta_{23} \end{array}
\left[ \begin{array}{ccc} 0 & \sqrt{A^2 + B^2} & 0 \\ \sqrt{A^2 + B^2} & 0
& 0 \\ 0 & 0 & 0 \end{array} \right] 
\begin{array}{c} \longrightarrow \\ \theta_{12} \end{array}
\left[ \begin{array}{ccc} \sqrt{A^2 + B^2} & 0 & 0 \\
0 & -\sqrt{A^2 + B^2} & 0 \\ 0 & 0 & 0 \end{array} \right].
\end{equation}

\noindent
Note that this sequence of large rotations is precisely Solution A 
of Puzzle 3. Even though the hierarchy of neutrino masses
is inverted here, in the sense that $m_3$ is the smallest, the near
degeneracy of $|m_1|$ and $|m_2|$ gives the correct hierarchy of splittings,
$\delta m^2_{sol} \ll \delta m^2_{atm}$, thus resolving Puzzle 2.

{\it Factorized Mass Matrix Models.} In some models $M_{\nu}$ has 
approximately the form 
\begin{equation}
M_{\nu} \cong \left( \begin{array}{ccc} 0 & c^2 & d^2 \\ c^2 & B^2 & AB \\
d^2 & AB & A^2 \end{array} \right),
\end{equation}

\noindent
where $c,d \ll A \sim B$. One can see that in a sense this form is
the opposite of the inverted hierarchy form. (In fact, if $M_{\nu}$
has this form, then $M_{\nu}^{-1}$ has the inverted hierarchy form.)
We call this form factorized, because if $m_{ij}$ is the 23 block
of $M_{\nu}$, then $m_{ij} \cong a_i a_j$. This form can arise
if the dominant contribution to $M_{\nu}$ comes from integrating out
a single heavy fermion $N$ that has Dirac mass $m_i(\nu_i N)$ with the
left-handed neutrinos $\nu_2$ and $\nu_3$ (its coupling to $\nu_1$
should be smaller). A notable instance of this occurs in the 
SUSY models with R-parity violation, where $N$ is the neutralino.

A rotation in the 23 plane by $\theta_{23} = \tan^{-1} (B/A) = O(1)$
eliminates the 23, 32, and 22 elements in Eq. (6). That leaves a matrix
whose 12 block has a ``pseudo-Dirac" form, with the 12 and 21 elements
being larger than the 11 and 22 elements. This block can be
diagonalized with a rotation in the 12 plane by $\theta_{12} \cong
\pi/4$. The resulting matrix can be diagonalized with only small
further rotations. Thus, as in the inverted hierarchy models, one has just
the sequence of large rotations corresponding to Solution A of Puzzle 3. 
The hierarchy of neutrino masses
is the ``normal" one with $m_1, m_2 \ll m_3$, giving the correct
hierarchy of splittings and resolving Puzzle 2.

{\it Flavor Democracy Models.} A third kind of model, about which
many papers have been written \cite{bd}, 
assumes that the $M_L$, $M_U$, and $M_D$
all have approximately the ``democratic" form in which all the entries are
equal. The matrix $M_{\nu}$, is assumed to be approximately diagonal.
The CKM angles thus end up being small due to cancellation, whereas
the large leptonic angles that come from diagonalizing $M_L$ do not get 
cancelled. The democratic form can be diagonalized by the sequence
of rotations 
\begin{equation}
\left( \begin{array}{ccc}
1 & 1 & 1 \\ 1 & 1 & 1 \\ 1 & 1 & 1 \end{array} \right) 
\begin{array}{c} \longrightarrow \\ \theta_{12} \end{array}
\left( \begin{array}{ccc} 0 & 0 & 0 \\ 0 & 2 & \sqrt{2} \\
0 & \sqrt{2} & 1 \end{array} \right)
\begin{array}{c} \longrightarrow \\ \theta_{23} \end{array}
\left( \begin{array}{ccc} 0 & 0 & 0 \\ 0 & 0 & 0 \\
0 & 0 & 3 \end{array} \right),
\end{equation}

\noindent
where $\theta_{12} = \tan^{-1} 1 = \pi/4$ and $\theta_{23} =
\tan^{-1} \sqrt{2}$. This sequence of large rotations is just that of 
Solution B of Puzzle 3. The flavor democracy
models also give the ``normal" neutrino mass hierarchy, resolving Puzzle
2.

In all three kinds of non-see-saw model we have discussed we see
that $\theta_{sol} \cong \pi/4$ (maximal mixing), whereas $\theta_{atm}$
is only predicted to be large, but not nearly maximal (though it may be
by accident). Curiously, the empirical situation is just the reverse.
It is $\theta_{atm}$ that is observed to be close to maximal. 
(The best-fit value is $\sin^2 2 \theta_{atm} = 1.0$.) This
is our fourth puzzle:

{\bf Puzzle 4:} {\it Why is $\theta_{atm}$ so close to maximal?}
It is not an accident that many models predict $\theta_{sol}$
to be nearly maximal while very few models \cite{mn} 
exist where $\theta_{atm}$ is.
The reason is essentially the following. The simplest way to arrange that
a mixing angle is nearly maximal is to assume that the relevant 
$2 \times 2$ block of the $3 \times 3$ mass matrix is pseudo-Dirac. 
The diagonalization of such a matrix leads to nearly degenerate masses, 
which is to say, very small $\delta m^2$.
For example, suppose one considers the matrix 

\begin{equation}
M_{\nu} \cong \left( \begin{array}{ccc} 0 & m & 0 \\ m & 0 & 0 \\
0 & 0 & m' \end{array} \right).
\end{equation}

\noindent
This gives $\theta_{sol} \cong \pi/4$, $m_1 \cong m_2 \cong m$, and
$m_3 \cong m'$. Thus one has $\delta m^2_{sol} \ll \delta m^2_{atm}$
as desired. Now consider the matrix

\begin{equation}
M_{\nu} \cong \left( \begin{array}{ccc} m' & 0 & 0 \\ 0 & 0 & m \\
0 & m & 0 \end{array} \right).
\end{equation}

\noindent
This gives $\theta_{atm} \cong \pi/4$, but now $m_3 \cong m_2 \cong m$
and $m_1 \cong m'$, so that $\delta m^2_{atm} \ll \delta m^2_{sol}$.
Thus, models with nearly maximal $\theta_{sol}$
tend to give $\delta m^2_{sol} \ll \delta m^2_{atm}$ as observed,
whereas with nearly maximal $\theta_{atm}$ will tend to give
the wrong result $\delta m^2_{atm} \ll \delta m^2_{sol}$.

\section{See-saw Models}

See-saw models have three great advantages over non-see-saw models.
First, they do not have to invent an exotic mechanism for generating
neutrino mass. There is nothing exotic about right-handed neutrinos,
which indeed have to exist in most kinds of gauge-unified models
($SU(5)$ being an exception). Indeed, grand unification, which is
well motivated on other grounds, naturally leads to see-saw neutrino masses.
Second, see-saw/GUT models beautifully explain the scale of neutrino mass.
Writing the heaviest neutrino mass as $m_3 = m_t^2/M_R$ (GUTs typically
relate the neutrino Dirac masses to the up-type quark masses), and
taking $m_3 \cong \sqrt{m_3^2 - m_2^2} = \sqrt{\delta m^2_{atm}} \cong
0.06$ eV,
one finds, $M_R \sim 10^{15}$ GeV, which is
very close to the GUT scale known from running of the gauge couplings.
By contrast, in non-see-saw models the neutrino mass scale depends on
many parameters about which virtually nothing is known even as to their
order of magnitude. Third, see-saw/GUT models tend to be far
more predictive than most non-see-saw schemes.

At first glance, Puzzle 1 seems especially puzzling in the context of
see-saw/GUT models, since grand unification closely relates quarks and
leptons. And it is certainly true that historically the great majority 
of GUT models predicted leptonic mixing angles of the same
order as the CKM angles. Looking more closely, however, we see that
this need not be the case. Indeed, there is a beautiful way to resolve
Puzzle 1 in the see-saw/GUT framework.

All grand unified gauge groups contain $SU(5)$ as a subgroup, and $SU(5)$
relates down-type quarks to charged leptons {\it having the opposite 
chirality}. The $\overline{{\bf 5}}$ contains $\ell^-_L$ and the
charge conjugate of $d_R$, while the ${\bf 10}$ contains $d_L$ and the
charge conjugate of $\ell^-_R$. As a consequence, what is related by
$SU(5)$ is $\theta_{d_L} \longleftrightarrow \theta_{\ell_R}$ and 
$\theta_{d_R} \longleftrightarrow \theta_{\ell_L}$. Since the CKM
angles are the mixings of {\it left-handed} quarks, and the MNS angles are
the mixings of {\it left-handed} leptons, $SU(5)$ does not relate them
to each other. Rather, it relates the CKM angles to some unobserved
mixing of right-handed leptons, and the MNS angles to some unobserved
mixing of right-handed quarks. Consequently, it is perfectly possible
for the CKM angles to be small and the ``corresponding" MNS angles large
if the mass matrices $M_L$ and $M_D$ are highly left-right asymmetric
or ``lopsided" \cite{lop}. We can see this in a simple toy model.

Consider an $SU(5)$ model with mass terms for the second and third family
given by $\lambda ({\bf 10}_3 \overline{{\bf 5}}_3 + 
\sigma {\bf 10}_3 \overline{{\bf 5}}_2
+ \epsilon {\bf 10}_2 \overline{{\bf 5}}_3) \langle \overline{{\bf 5}}_H
\rangle$, with $\epsilon \ll \sigma \sim 1$. The mass matrices $M_L$ and 
$M_D$ appear in the terms $\overline{\ell_R}_i (M_L)_{ij} \ell_{Lj}
+ \overline{d_R}_i (M_D)_{ij} d_{Lj}$, so they have the form
\begin{equation}
M_L = \left( \begin{array}{ccc} - & - & - \\ - & 0 & \epsilon \\ 
- & \sigma & 1 \end{array}
\right) m, \;\;\; M_D = \left( \begin{array}{ccc} - & - & -  \\
- & 0 & \sigma \\ - & \epsilon & 1 \end{array}
\right) m,
\end{equation}

\noindent
where the dashes are small entries for the first family coming from other 
terms. Note the left-right transposition between $M_L$ and $M_D$, whose origin 
we have already explained. The 32 entry in $M_L$ is $\sigma$, which gives
an $O(1)$ contribution to the MNS angle $U_{\mu 3}$. The 32 entry
in $M_D$, on the other hand, is small and gives only a small
contribution to the CKM angle $V_{ub}$. (The 23 entries control
mixings of right-handed fermions.)

There is an interesting feature of the quark and lepton mixings that
is explained very elegantly by such lopsided ``textures" as in Eq. (10).
Many models
are based on symmetric ``textures" that are extensions of the $2 \times 2$
matrix shown in Eq. (1). As we saw, such textures tend to relate
the mixing angles to the square-roots of mass ratios. A typical 
prediction for the quarks is

\begin{equation}
V_{cb} = \sqrt{|m_s/m_b|} - e^{i \phi} \sqrt{|m_c/m_t|}.
\end{equation}

\noindent
The first term on the right-hand side is about $0.14$, and the second
about $0.05$, whereas $V_{cb} \cong 0.04$, so that the prediction for 
$V_{cb}$ of such symmetric-texture models tends to be about a factor 
of 2 or 3 too large. The analogous relation for leptons is

\begin{equation}
U_{\mu 3} = \sqrt{|m_{\mu}/m_{\tau}|} - e^{i \phi'} \sqrt{|m_2/
m_3|}.
\end{equation}

\noindent
Here the first term on the right-hand side is $0.24$, and the second
less than about $0.1$ (assuming hierarchical neutrino masses, so
that $m_3 \cong \sqrt{\delta m^2_{atm}}$ and $m_2 \cong 
\sqrt{\delta m^2_{sol}}$), whereas $U_{\mu 3} \cong 0.7$.
Thus the prediction for $U_{\mu 3}$ in such symmetric-texture models
tends to be about a factor of 2 or 3 too small. That symmetric
textures give $V_{cb}$ too large and $U_{\mu 3}$ too small by about
the same factor is readily explained by the assumption that the textures
are in reality not symmetric but lopsided. 
We see from Eq. (10) that for the
lopsided textures the mass ratios of second family to third family
fermions are of order $\sigma \epsilon$. (The third eigenvalue is
$\cong m$ while the product of the second and third eigenvalues
is just the determinant of the 23 block or $- \sigma \epsilon m^2$.)
Since $V_{cb} \sim \epsilon$ and $U_{\mu 3} \sim \sigma$, one expects
$V_{cb} \sim \sqrt{\epsilon/\sigma} \sqrt{m_s/m_b}$ and 
$U_{\mu 3} \sim \sqrt{\sigma/\epsilon} \sqrt{m_{\mu}/m_{\tau}}$. This is
just what is observed if $\sqrt{\sigma/\epsilon} \sim 2$ or $3$.
In other words, in lopsided models the smallness of $V_{cb}$ and
the largeness of $U_{\mu 3}$ are seen to be two sides of the same coin.
(In realistic lopsided models the textures are similar in form but not
exactly the same as Eq. (10); however, the qualitative argument just given 
still applies.)

I said that see-saw/GUT models are in general more predictive than 
non-see-saw models. And, indeed, simple and highly predictive $SO(10)$ 
models that are very similar (for the second and third families) to the 
toy model just described have been constructed \cite{ab-bpw}. 
In fact, many models
based on lopsided mass matrices now exist in the literature \cite{bd}.

Note the very important point that in such lopsided models the large
atmospheric neutrino angle comes from the {\it charged lepton mass matrix}
$M_L$ rather than from $M_{\nu}$. This shows how such models resolve Puzzle
2. In lopsided models the reason why some of the neutrino mixing angles
can be large even though all the neutrino mass ratios are small is that
large neutrino mixing angles can be caused by large off-diagonal elements
in $M_L$ (here $\sigma$) whereas the neutrino mass ratios obviously are 
determined entirely by $M_{\nu}$.

How can lopsided models resolve Puzzle 3? There are two interesting
and simple possibilities. One possibility is that the large $\theta_{atm}$
arises from $M_L$ as just described, but that the large $\theta_{sol}$
arises from $M_{\nu}$. This corresponds to Solution C. Such models
are very easy to construct \cite{ab-bimax}. 
The other possibility is that both of the
large angles $\theta_{atm}$ and $\theta_{sol}$ come from lopsidedness
in $M_L$ \cite{bb-bimax}. Consider the following matrix
\begin{equation}
M_L = \left( \begin{array}{ccc} - & - & - \\ - & - & \epsilon \\
\rho' & \rho & 1 \end{array} \right) m,
\end{equation}

\noindent
where $\epsilon \ll \rho' \sim \rho \sim 1$ and the dashes represent
elements yet smaller than $\epsilon$. A rotation in the 12 plane by
$\theta_{atm} \cong \tan^{-1} (\rho'/\rho)$ 
brings the matrix to the form shown in Eq. (8) with
$\sigma = \sqrt{\rho^2 + \rho^{\prime 2}}$. Then a rotation in the 23 plane
by angle $\theta_{23} \cong \tan^{-1} \sigma$, as in the toy model, 
eliminates the large 32 element. The further rotations required to
diagonalize $M_L$ are small. This sequence of large rotations
in the charged lepton sector gives just Solution B of Puzzle 3. 
(It should be noted that, as in Eq. (10),
the matrix $M_D$ will have the large elements appearing transposed 
compared to $M_L$, so that they only affect mixings of right-handed
quarks.)

Very few models in the literature attempt to explain why
$\theta_{atm}$ is nearly maximal (Puzzle 4). It turns out that within the
framework of lopsided models it is not difficult to obtain
$\theta_{atm} \cong \pi/4$ \cite{bb-bimax}. 
Consider a model with $M_L$ having the
form in Eq. (9), where some nonabelian symmetry relates $\mu^-_L$
and $\tau^-_L$ so that $\rho = 1$. That would give the relations
$\tan \theta_{sol} \cong \rho'$ and $\tan \theta_{atm} \cong \sqrt{1 + 
\rho^{\prime 2}}$, which imply the interesting relation
$\tan^2 \theta_{atm} \cong 1 + \tan^2 \theta_{sol}$, or
equivalently $\sin^2 2 \theta_{atm} \cong (1 + \tan^2 \theta_{sol})/
(1 + \frac{1}{2} \tan^2 \theta_{sol})^2$. Even for the
best-fit LMA value $\tan^2 \theta_{sol} \approx 0.4$ this gives
$\sin^2 2 \theta_{atm} \cong 0.97.$


\begin{thebibliography}{99}

\bibitem{wzf} S. Weinberg, Trans. N.Y. Acad. Sci. {\bf 38}, 185 (1977);
F. Wilczek and A. Zee, {\it Phys. Lett.} {\bf B70}, 418 (1977); 
H. Fritzsch, {\it Phys. Lett.} {\bf B70}, 436 (1977).

\bibitem{bn} K.S. Babu and S. Nandi, {\it Phys. Rev.} {\bf D62}, 033002 
(2000). 

\bibitem{mns} Z. Maki, M. Nakagawa, and S. Sakata, {\it Prog.
Theor. Phys.} {\bf 28}, 870 (1962).

\bibitem{fits} J.N. Bahcall, M.C. Gonzalez-Garcia, C. Pena-Garay,
{\it JHEP} {\bf 0108}, 014 (2001); hep-ph/0204314.

\bibitem{random} L.J. Hall, H. Murayama, and N. Weiner, {\it Phys. Rev.
Lett.} {\bf 84}, 2572 (2000).

\bibitem{bd} For a review of models and further references see 
S.M. Barr and I. Dorsner, {\it Nucl. Phys.} {\bf B585}, 79 (2000).

\bibitem{zee} A. Zee, {\it Phys. Lett.} {\bf B93}, 389 (1980);
{\it ibid.} {\bf B161}, 141 (1985).

\bibitem{mn} R.N. Mohapatra and S. Nussinov, {\it Phys. Rev.} {\bf D60},
013002 (1999).

\bibitem{lop} K.S. Babu and S.M. Barr, {\it Phys. Lett.} {\bf B381},
202 (1996); J. Sato and T. Yanagida, {\it Phys. Lett.} {\bf B430}, 127
(1998); C.H. Albright, K.S. Babu and S.M. Barr, {\it Phys. Rev. Lett}
{\bf 81}, 1167 (1998); N. Irges, S. Lavignac, and P. Ramond,
{\it Phys. Rev.} {\bf D58}, 035003 (1998).

\bibitem{ab-bpw} C.H. Albright and S.M. Barr, {\it Phys. Lett.} {\bf B452}
287 (1999); 

\bibitem{ab-bimax} C.H. Albright and S.M. Barr, {\it Phys. Lett.} {\bf B461}
218 (1999). 

\bibitem{bb-bimax} K.S. Babu and S.M. Barr, {\it Phys. Lett.} {\bf B525},
289 (2002).

\end{thebibliography}
\end{document}